\documentclass[12pt]{article}
\usepackage{fullpage}
\usepackage{epsfig}
\usepackage{amsmath}
\usepackage{amssymb}
\usepackage{color}

\begin{document}


\centerline{\bf\Large Scalar mesons on the Lattice\footnote{Talk presented at 
{\it Mini-workshop  Exciting Hadrons} at Bled, Slovenia, July 2005.}}

\vspace{1cm}

\centerline{\large Sasa Prelovsek\footnote{Electronic address: {\it sasa.prelovsek@ijs.si}}}

\vspace{0.5cm} 

\centerline{\small \it Department of Physics, University of Ljubljana, Jadranska 19,  1000 Ljubljana, Slovenia}

\centerline{\small \it and}

\centerline{\small \it Institute Jozef Stefan, Jamova 39,  1000 Ljubljana, Slovenia}

\vspace{1cm} 

\centerline{\bf Abstract}

\vspace{0.2cm}

The simulations of the light scalar mesons on the lattice are presented at 
the introductory level. The methods for determining 
the scalar meson masses are described. The problems related to
 some of these methods are presented and their solutions discussed.

\vspace{0.8cm}

\section{Introduction}

\begin{figure}[htb!]
\begin{center}
\epsfig{file=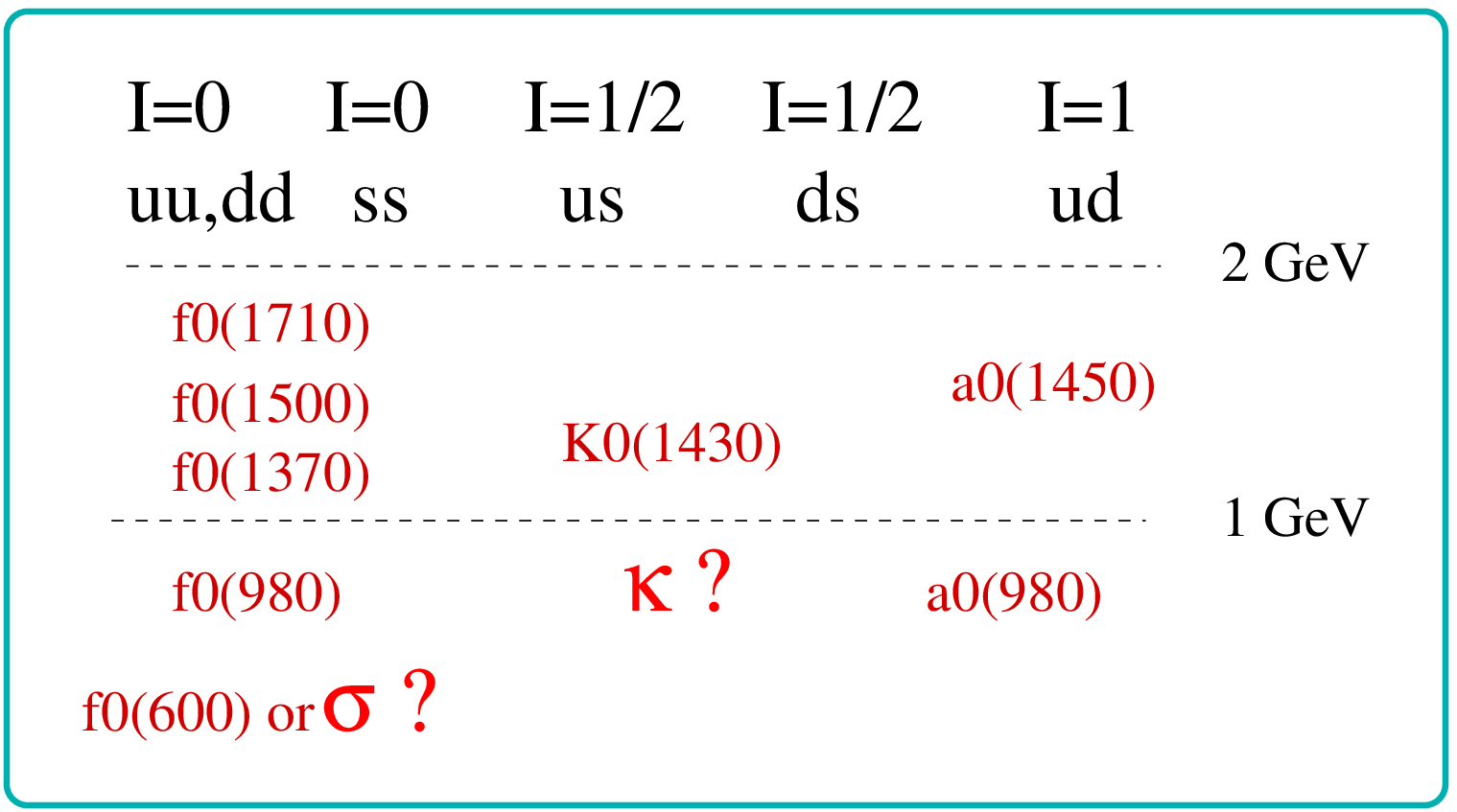,height=5cm}
\end{center}

\vspace{-0.8cm}

\caption{ \small The spectrum of observed light scalar resonances below 
$2$ GeV \cite{pdg}. The existence of $\sigma$ and $\kappa$ are still very controversial 
experimentally. }\label{fig.spectrum}
\end{figure}

The observed spectrum of the light scalar resonances below $2$ GeV 
is shown in Fig. \ref{fig.spectrum}. The existence of 
flavor singlet $\sigma$ and strange iso-doublet $\kappa$ 
are still very controversial \cite{pdg}. Irrespective of their existence, 
it is difficult to describe all the observed resonances 
by one or two $SU(3)$ flavor nontes of $\bar qq$ states:

\vspace{-0.2cm}

\begin{itemize}
\item
If $\sigma$ and $\kappa$ do not exist, than $K_0(1430)$ has to be strange 
partner of $a_0(980)$, but the mass difference appears to big. 
Also there are to many states to be described by one nonet.

\vspace{-0.2cm}

\item 
If $\sigma$ and $\kappa$ exist, then all these states could represent two 
$\bar qq$ nonets and one glueball, where the largest glueball component 
is commonly attributed to $f_0(1500)$. However, most of the models and 
lattice simulations have difficulties in relating the observed properties 
of states below $1$ GeV to the $\bar qq$ states. 
\end{itemize}

\vspace{-0.2cm}

This situation is in contrast to the spectrum of light pseudoscalar, 
vector and 
axial-vector resonances, where $\bar qq$ assignment works well.  
It raises a question whether the scalar resonances 
 below $1$ GeV are conventional 
$\bar qq$ states or perhaps exotic 
states such as tetraquarks \cite{tetraquark}. 

This issues could be settled if the mass of the lightest $\bar qq$ states 
could be reliably determined on the lattice and identified with the 
observed resonances. In lattice QCD, the 
 hadron masses are conventionally extracted from 
the correlation functions that are computed on the discretized space-time. 

In the next section we present how the scalar correlator is calculated 
on the lattice. The relation between the scalar 
correlator  and the scalar 
meson mass is derived in Section 3.  A result for the 
 mass of $I=1$ scalar meson 
is presented in Section 4. In Section 5 we point out the problems 
which arise due to 
the unphysical approximations that are often used in the lattice simulations  
and we discuss the proposed solutions.
 We close with Conclusions. 

This article follows the introductory spirit of the talk given at the 
Workshop {\it Exciting hadrons}$^1$ and many technical details are omitted.

\section{Calculation of the scalar correlator}

\begin{figure}[htb!]
\begin{center}
\epsfig{file=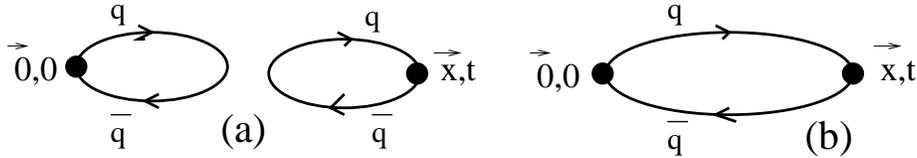,height=2.5cm}
\end{center}

\vspace{-0.8cm}

\caption{ \small The disconnected (a) and the connected (b) Feynman diagrams 
that need to be evaluated to compute 
the correlator. The disconnected diagram is  
present only for the flavor singlet meson.  }\label{fig.quark}
\end{figure}

Let us consider the correlation function for a {\it flavor non-singlet 
scalar meson} $\bar q_1q_2$ first. In a lattice simulation 
it is calculated using the Feynman functional integral 
on a discretized space-time of finite volume and 
finite lattice spacing.   
The correlation function represents a creation 
of a pair  $\bar q_1q_2$ with $J^P=0^+$ at time zero and annihilation of 
the  same pair at some later Euclidean time $t$
\begin{equation}
\label{cor_0}
C(t)=\sum_{\vec x}\langle 0|\bar q_1(\vec x,t)q_2(\vec x,t)~\bar q_2(\vec 0,0)q_1(\vec 0 ,0)|0 \rangle~,
\end{equation}
where both quarks are created (annihilated) at the same spatial point 
for definiteness here\footnote{Different shapes of creation and annihilation 
operators in spatial direction can be used.}. 
Wick contraction relates this to the product of two quark propagators 
shown by the connected diagram in Fig. \ref{fig.quark}b
\begin{align}
\label{cor_1}
C(t)&=\bigl\langle C_G(t)\bigr\rangle_{G} \\
C_G(t)&= \sum_{\vec x}  {\rm Tr}_{s,c} \bigl[ {\rm Prop}^2_{\vec 0,0\to\vec x, t}~{\rm Prop}^1_{\vec x,t\to\vec 0,0}\bigr]
\!=\sum_{\vec x}  {\rm Tr}_{s,c} \bigl[ {\rm Prop}^2_{\vec 0,0\to\vec x, t}\gamma_5 {\rm Prop}^{1~\dagger}_{\vec 0,0\to \vec x,t}\gamma_5\bigr]\nonumber ~.
\end{align}
The quark propagator in the gluon field $G$ and Euclidean space-time \cite{roethe} 
\begin{equation}
\label{prop}
{\rm Prop^i}_{\vec x,x_0\to\vec y,y_0 }=\biggl(\frac{1}{{\not{\! \! D_E}}+m_{i}}\biggr)_{\vec x,x_0\to\vec y, y_0}
\end{equation} 
is the inverse of the discretized Dirac operator   
${\not{\! \! D_E}}+m_{i}~$, which is a matrix in coordinate space and 
 depends on the gluon field $G$ \footnote{
${\not{\!\! D}}=\gamma^\mu (\partial_\mu+\tfrac{i}{2}\lambda_a G^a_\mu)$ 
in continuum Minkowski space-time. }. 
The inversion of a large Dirac matrix is numerically costly, 
but the calculation of 
correlator (\ref{cor_1}) is feasible since it depends only on two propagators 
from a certain point $(\vec 0,0)$ 
to all points $(\vec x,t)$. Both of these are obtained by solving 
the equation $({\not{\! \! D_E}}+m_i)V^\prime=V$ 
for a single\footnote{In fact $({\not{ \!\! D_E}}+m_i)V^\prime=V$ 
has to be solved for every spin and color 
of the source vector $V$.}  source vector $V$ which is non-zero only at 
$(\vec 0,0)$.  The expectation value over the gluon fields in (\ref{cor_1}) 
is computed based on the Feynman functional integral 
\begin{equation}
\label{cor_2}
C(t)=\frac{\int {\cal D}G ~C_G(t)~\int {\cal D}q\int {\cal D}\bar q~e^{-S_{QCD}}}{\int {\cal D}G\int {\cal D}q\int {\cal D}\bar q~e^{-S_{QCD}}}=\frac{\int {\cal D}G ~C_G(t)~\Pi_i {\rm det}[{\not{\!\! D_E}}+m_i]~e^{-S_{G}}}{\int {\cal D}G~\Pi_i {\rm det}[{\not{\!\! D_E}}+m_i]~e^{-S_{G}}}~.
\end{equation}
A finite ensemble of $N$ gluon field 
configurations is generated in the lattice simulations.  
Each configuration is generated with a probability 
$\Pi_i {\rm det}[{\not{\!\! D_E}}+m_i]~e^{-S_{G}}$ for a given discretized 
gauge action $S_G$ and Dirac operator ${\not{\!\! D_E}}$. 
The functional integral (\ref{cor_2}) 
is calculated as a sum over the ensemble 
\begin{equation}
C(t)=\frac{1}{N}\sum_{j=1}^{N}C_{G_j}(t)~.
\end{equation}

The correlator for the {\it flavor singlet scalar meson} $\bar qq$
\begin{equation}
\label{cor_3}
C(t)=\sum_{\vec x}\langle 0|\bar q(\vec x,t)q(\vec x,t)~\bar q(\vec 0,0)q(\vec 0 ,0)|0 \rangle~
\end{equation}
requires  also the calculation of the 
disconnected diagram in Fig. \ref{fig.quark}a 
\begin{equation}
\label{cor_4}
\biggl \langle {\rm Tr}_{s,c}{\rm Prop}_{\vec 0,0\to\vec 0, 0}~\sum_{\vec x}{\rm Tr}_{s,c}{\rm Prop}_{\vec x,t\to\vec x,t}\biggr\rangle_G
\end{equation}
in addition to connected one. The propagator ${\rm Prop}_{\vec x,t\to\vec x,t}$
in principle requires the solution of $({\not{ \!\! D_E}}+m_i)V^\prime=V$ 
for source vector $V$ at any point. Such a number of inversions 
is normally prohibitively large and one is forced 
to use approximate methods for evaluating the disconnected part (\ref{cor_4}) 
of the singlet correlator. A calculation of the correlator for  singlet 
meson in therefore much more demanding than for 
non-singlet meson. 

\section{Relation between correlator and meson mass}

In this Section we derive the relation between the scalar correlator  
and the scalar meson mass. The state 
$\bar q(\vec 0)q(\vec 0)|0\rangle$ that is created 
at time zero is not a scalar meson $|S\rangle$, 
but it is a superposition of the scalar meson  and 
all the other eigenstates of Hamiltonian $|n\rangle$ with the 
same quantum numbers  $J^P=0^+$ and $I^G$ as $|\bar qq\rangle$ 
\begin{equation}
\label{sum}
|\bar qq\rangle=\sum_n c_n|n\rangle =c_1 |S\rangle+c_2 |S^*\rangle+
\sum c_i \biggl \vert{\rm\small {multi\atop hadron\ st.}}\biggr\rangle_i+...\biggl( +c_0|0\rangle\ {\rm \small{only\ for \atop singlet}}\biggr )~.
\end{equation}
Here $|S\rangle$ and $|S^*\rangle $ are ground and excited  
scalar mesons, while the third term represents the sum over 
 multi-hadron states. 
The eigenstate $|n\rangle$ evolves as $e^{i\vec p_n\vec x-E_nt}$ 
in Euclidean space-time, 
so the scalar correlators (\ref{cor_0}) and (\ref{cor_3}) evolve as
\begin{align}
\label{mass_fit}
C(t)&=\!\!\sum_{\vec x}\langle \bar q(\vec x,t)q(\vec x,t)~\bar q(\vec 0,0)q(\vec 0 ,0)\rangle=\!\!\sum_n \sum_{\vec x}\langle \bar qq|n\rangle e^{i\vec p_n\vec x-E_nt}\langle n|\bar qq\rangle=\!\!\sum_n |\langle \bar qq|n\rangle|^2 ~e^{-E_nt}\bigl\vert_{\vec p=0}
\nonumber\\
&=|c_1|^2e^{-m_S t}+|c_2|^2e^{-m_{S^*} t}+\sum |c_i|^2e^{-E^{\rm {multi\atop had.}}_i t}+ ... \biggl (+|c_0|^2 \ {\rm \small{only\ for \atop singlet}}\biggr )~. \end{align}

If $|S\rangle$ is the lightest state among $|n\rangle$, 
than $C(t)\propto e^{-m_St}$ at large $t$ 
and $m_S$ and can be extracted simply by fitting the lattice correlator 
to the exponential time dependence. 

In the case of the {\it flavor singlet} correlator, 
the lightest state in the sum 
(\ref{sum}) is the vacuum state. Its corresponding coefficient 
$c_0$ (\ref{sum}) is the scalar condensate $\langle \bar qq\rangle$. 
Another important light state that contributes at large $t$ is $\pi\pi$,  
so extraction of $m_\sigma$ requires the fit to 
\begin{equation}
\label{cor_sigma}
C(t)\stackrel{t\to\infty}{=}|c_\sigma|^2e^{-m_\sigma t}+
\sum_{\vec p_{\pi}} \bigl\vert c_{\vec p_\pi}\bigr\vert^2 e^{-E^{\pi\pi}_{\vec p_\pi}t}+\langle \bar qq\rangle^2~.
\end{equation}
The extraction of $m_\sigma$ is very 
challenging since $C(t)$  requires the calculation 
of the disconnected diagram (see previous Section) and since RHS 
in (\ref{cor_sigma}) 
is largely dominated by $\langle \bar qq\rangle^2$. 
 
These two problems do not affect the
 study of the {\it flavor non-singlet} meson.  
However, even in this case there are several multi-hadron states 
which are light and need to be taken into account in the fit  
 of the correlator (\ref{mass_fit}) at large $t$ 
in order to extract $m_S$. The lightest multi-hadrons states 
with $J^P=0^+$ are two-pseudoscalar states in $S$-wave.  
In case of $I=1$ correlator, the contribution 
of scalar meson $a_0$ is accompanied by contributions of 
$\pi\eta$, $\bar KK$  and $\pi \eta^\prime$ in three-flavor QCD. 
Let us note that in nature  
 these three states are  lighter than observed resonance $a_0(1450)$; 
the state $\pi\eta$ is also lighter than observed resonance $a_0(980)$.   
In two-flavor QCD,  the only two-pseudoscalar state $\pi\eta^\prime$ 
is relatively heavy and not so disturbing for the extraction of 
$m_{a0}$ from (\ref{mass_fit}).

\vspace{0.2cm}

The above derivation of time-dependence for a correlator 
was based on QCD, which is a proper unitary field theory. The resulting 
correlator (\ref{mass_fit}) is positive definite. Let us point out 
that certain approximations used in lattice simulations 
(quenching, partial quenching, staggered fermions, mixed-quark actions) 
break unitarity and 
may render negative correlation function. 
These approximations will be discussed in Section 5 together with 
the necessary modifications of the fitting formula (\ref{mass_fit}). 

\section{Mass of scalar meson with I=1}

A lattice simulation of the scalar meson $a_0$ with $I=1$
\cite{sasa_pq} is presented in 
this section, as an example. It employs   
two dynamical quarks\footnote{Fermion 
determinant in (\ref{cor_2}) incorporates quarks $i=u,d$.}, lattice spacing 
$0.12$ fm, lattice volume $16^3\times 32$ and ensemble of about 
$100$ gauge configurations \cite{sasa_pq,dyn_dwf}. 
The advantage of simulation \cite{sasa_pq} is that its discretized 
(Domain-Wall) fermion action has good chiral properties:
it is invariant under the chiral transformation for $m_q=0$ even at finite 
lattice spacing\footnote{This is strictly true only when the 5th dimension 
in Domain-Wall fermion action is infinitely large.}, which is not the case for some of the  
commonly used discretized fermion actions. 
 Another advantage of the simulation with two dynamical 
quarks 
\cite{sasa_pq} is that the exponential fit of the correlator at large $t$ 
renders $m_{a0}$. The conventional exponential 
fit is justified in this case since the only two-pseudoscalar 
intermediate state   in two-flavor QCD is $\pi\eta^\prime$, 
which is relatively heavy and does not affect the extraction 
of $m_{a0}$ (see previous Section). 

\begin{figure}[htb!]
\begin{center}
\epsfig{file=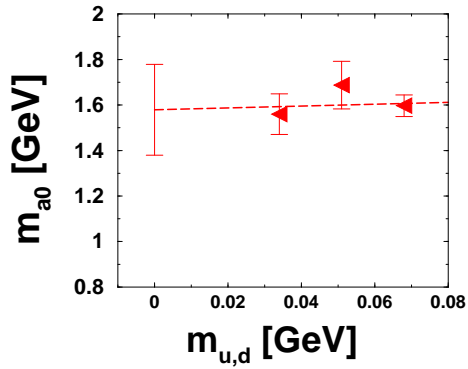,height=5cm}
\end{center}

\vspace{-0.8cm}

\caption{ \small The triangles present resulting $m_{a0}$ for three values of bare quark masses $m_{u,d}$ \cite{sasa_pq}. The dashed line is the 
linear extrapolation of $m_{a0}$ 
to the value of $m_{u,d}$ in nature. }\label{fig.m_dyn}
\end{figure}

The resulting mass is presented in Fig. \ref{fig.m_dyn} for different 
input masses $m_{u,d}$, where isospin limit $m_u=m_d$ is employed. 
There are no simulations at physical masses $m_{u,d}$ since the 
pion cloud around the scalar meson with $\lambda_\pi=hc/140~$MeV$\simeq 9$ fm
   would be to squeezed 
on the lattice with extent $16\times 0.12~$fm$\simeq 2~$fm. 
The $u/d$ quarks and pions are heavier in simulation than in the nature 
in order to avoid large finite volume effects. The linear extrapolation 
of $m_{a0}$ to the physical quark mass $m_{u,d}\simeq 4~{\rm MeV}$ 
in Fig.  \ref{fig.m_dyn} gives 
\begin{equation}
\label{m_a0}
m_{a0}=1.58\pm 0.34~{\rm GeV}~.
\end{equation}
Although our result for the mass of the lightest $\bar qq$ state with $I=1$ 
 has sizable error-bar, it appears to  
be closer to the observed resonance $a_0(1450)$ than to $a_0(980)$. 
It gives preference to the interpretation that $a_0(980)$ is not 
conventional $\bar qq$ state. 

Results from other lattice  simulations of the light scalar mesons 
can be found in \cite{bardeen}-\cite{staggered}.

\section{Problems due to unphysical approximations}

The simulation presented in the previous section 
is a discretized version of two-flavor QCD 
and does not employ any unphysical approximations 
except for  the discretization of space-time. 
It renders positive definite correlation function, as expected in proper 
Quantum Field Theory (\ref{mass_fit}). 

However, lattice simulations often employ unphysical approximations 
which facilitate numerical evaluation. One of the indications 
that the simulation does not correspond to a proper QCD is 
the negative scalar correlator. Another sign of unphysical simulation  is 
when  $I=1$ 
correlator drops as $e^{-2M_\pi t}$ at large $t$ although 
the lightest two-pseudoscalar state with $I=1$ is $\pi\eta$. 
 Both of these unphysical lattice results can occur if the theory 
that is being simulated is not unitary, which is the case 
for all the commonly used approximations listed below:

\vspace{-0.1cm}

\begin{itemize}
\item In {\it quenched} simulation the fermion determinant in (\ref{cor_2}) 
is replaced by a constant. This corresponds to neglecting all the 
closed sea-quark loops. The $I=1$ scalar correlator 
is negative in this case  and its  
 negativity was attributed to the 
intermediate state $\pi\eta^\prime$  
 in Ref.
 \cite{bardeen}. The prediction for $\pi\eta^\prime$ intermediate state 
in quenched version of Chiral Perturbation Theory (ChPT) 
describes the sign and the magnitude of the lattice correlator at large $t$ 
well \cite{bardeen,sasa_quenched}. The  mass $m_{a0}$ was  extracted 
\cite{bardeen,sasa_quenched}
by fitting the quenched $I=1$ correlator to the sum of $e^{-m_{a0}t}$ 
term and 
 the contribution of $\pi\eta^\prime$ as predicted by Quenched ChPT.

\vspace{-0.1cm}

\item   In {\it partially quenched} simulation the mass of the sea quark is 
different from the mass of the valence quark, 
although they are the same in nature. 
The mass of the valence quark is the mass that appears in the propagator of the correlator $C_G$ (\ref{cor_1}), while  
the mass of the sea quark is the mass that appears in the fermion determinant 
(\ref{cor_2}). 
The partially quenched 
scalar correlator with $I=1$ was found to be negative if $m_{val}<m_{sea}$ 
\cite{sasa_pq}. This was attributed to intermediate states with 
two pseudoscalar mesons and was described well using 
partially quenched version of ChPT \cite{sasa_pq}. 
The  mass $m_{a0}$ was  extracted 
by fitting the partially quenched correlator 
to the sum of $e^{-m_{a0}t}$ term and 
 the contribution of two-pseudoscalar states as predicted by 
Partially Quenched ChPT \cite{sasa_pq}. The resulting mass 
 agrees with the mass  (\ref{m_a0}). 

\vspace{-0.1cm}
  
\item The simulations with {\it mixed quark actions} employ different 
discretizations of the Dirac operator for valence and sea quarks. 
The method of extracting scalar meson mass from a such simulations was proposed  in \cite{sasa_staggered,mchpt_taku}.

\vspace{-0.1cm}

\item The simulations with {\it staggered quarks} use an artificial 
taste degree of freedom for quarks in order to solve fermion doubling 
problem \cite{roethe}. The method of extracting scalar meson mass from simulations with staggered quarks \cite{staggered} was proposed  in \cite{sasa_staggered}. 
\end{itemize}

\vspace{-0.1cm}

All these approximations modify the contribution of two-pseudoscalar 
intermediate states with respect to QCD. The effects of these approximations 
can be therefore determined by predicting the two-pseudoscalar contributions 
using appropriate versions of ChPT. These analytic predictions  
\cite{bardeen,sasa_pq,sasa_staggered,mchpt_taku} 
allow the extraction of the scalar meson mass from the correlator as long as the 
contribution of two-pseudoscalar intermediate states does not completely 
dominate over the $e^{-m_{S}t}$ term.  

\section{Conclusions}

The nature of scalar resonances below $1$ GeV is not established yet. 
A lattice determination of the masses for  ground  $\bar qq$ scalar 
states would  help to resolve the problem. 

In principle, the scalar mass can be extracted from the scalar correlator 
that is computed on the lattice. However, 
the interesting term $e^{-m_{S}t}$ in the correlator is accompanied by the 
contribution of two-pseudoscalar  states $e^{-E_{PP}t}$. 
The problem is that the energy of two-pseudoscalar states is small, 
so they 
 may dominate the correlator and complicate the extraction of scalar 
meson mass. On top of that, the contribution of two-pseudoscalar states 
is significantly affected by the unphysical approximations that are often used in lattice simulations. Luckily, these effects can be predicted using 
appropriate versions of Chiral Perturbation Theory and they agree with the 
observed effects on the lattice correlators. We give the list of 
references, which provide the expressions for extracting $m_S$ from 
the correlators for various types of simulations. 

A simulation, which does not suffer from the problems listed above, 
gives $1.58\pm 0.34~$GeV for the mass of the lightest  
$\bar qq$ state with $I=1$.  This supports the interpretation that 
observed $a0(1450)$ is the lightest $(\bar qq)_{I=1}$ state, while 
$a0(980)$ might be something more exotic.

\end{document}